# Tuning of Graphene Properties via Controlled Exposure to Electron Beams

G. Liu, D. Teweldebrhan and A.A. Balandin

*Abstract*— **Controlled modification of graphene properties is essential for its proposed electronic applications. Here we describe a possibility of tuning electrical properties of graphene via electron beam irradiation. We show that by controlling the irradiation dose one can change the carrier mobility and increase the resistance at the minimum conduction point in the single layer graphene. The bilayer graphene is less susceptible to the electron beam irradiation. The modification of graphene properties via irradiation can be monitored and quantified by the changes in the disorder D peak in Raman spectrum of graphene. The obtained results may lead to a new method of** *defect engineering* **of graphene physical properties, and to the procedure of "writing" graphene circuits via e-beam irradiation. The results also have implications for fabrication of graphene nanodevices, which involve scanning electron microscopy and electron beam lithography.**

*Index Terms*— electron beam irradiation, graphene devices, Raman spectroscopy, disordered graphene, defects in graphene

## I. INTRODUCTION

GRAPHENE is a single sheet of $sp^2$-bound carbon atoms with many unique properties. It reveals extraordinary high room temperature (RT) carrier mobility of up to ~15,000 $cm^2$/Vs [1-3] and an extremely high "intrinsic" thermal conductivity exceeding ~3000 W/mK near RT for large flakes [4-6]. Recent experiments with modification of graphene surface via hydrogenation [7-8], potassium doping [9], ions irradiation [10] and adsorption of individual gas molecules ($NO_2$, $NH_3$, etc.) [11] have shown that graphene's properties can be altered and tuned for specific applications. However, little is known about the effect of the electron beam irradiation on graphene or graphene-based devices. The focused beams of electrons, which are commonly used in scanning electron microscopy (SEM) and device fabrication, are known to induce changes to the properties of carbon allotropes and nanostructures including graphite, fullerene and carbon nanotubes [12].

G. Liu, D. Teweldebrhan, and A. A. Balandin are with the Nano-Device Laboratory, Department of Electrical Engineering and Materials Science and Engineering Program, Bourns College of Engineering, University of California at Riverside, Riverside, CA 92521 USA. The work at UCR was supported, in part, by DARPA – SRC through the FCRP Center on Functional Engineered Nano Architectonics (FENA). E-mail: balandin@ee.ucr.edu

Recently, it was also shown that graphene exposure to the electron beams (e-beams) results in modification of its surface [13-14]. We have demonstrated that electron irradiation leads to the appearance of the disorder D peak at ~1350 $cm^{-1}$ in the Raman spectra of irradiated graphene [13].

In this paper we report how electrical properties of the single layer graphene (SLG) depend on the irradiation dose, and correlate the current – voltage characteristics with the evolution of Raman spectrum of irradiated graphene. We also investigate the response of bilayer graphene (BLG) on the electron beam irradiation and compare it with that of SLG. It is known that BLG reveals a band gap when subjected to electrical field and as a material might be more promising for electronic applications [3]. Our finding that BLG is less susceptible to electron beam irradiation, conventionally used in SEM characterization and device fabrication, adds extra motivation for the BLG device applications.

## II. FABRICATION AND MEASUREMENTS

The graphene flakes were prepared by the standard micro-mechanical exfoliation from the high quality graphite. The flakes were transferred to the silicon substrate with 300-nm-thick layer of silicon oxide. The Raman spectroscopy was used to verify the number of layers and check their quality. The details of our Raman inspection procedures were reported by us elsewhere [15-18]. The SLG and BLG samples were selected via de-convolution of the Raman 2D band and comparison of the intensities of the G peak and 2D band. The graphene back gate devices were fabricated with the electron beam lithography (EBL). We defined the source and drain regions and then followed with evaporation of Cr/Au with thickness of 10 nm and 60 nm respectively. The heavily doped silicon substrate was used as the back gate to tune the Fermi level of graphene.

We conducted electron beam irradiation using Leo SUPRA 55 electron-beam lithography (EBL) system, which allows for accurate control of the exposed area and irradiation dose. Special precautions have been taken to avoid additional unintentional e-beam irradiation. The alignment program in the utilized EBL system offers a way to scan only the alignment marks without exposing other locations. We used the gold alignment marks located more than 30 μm away from the graphene device to avoid unintentional irradiation during the scanning steps. For our experiments we selected the accelerating voltage of 20 keV and the working distance of 6 mm (the same as in EBL process). The area dosage was calculated and controlled by the nanometer pattern generation system (NPGS). NPGS allowed us to control the scanning distance from point to point and set the dwelling time on each point. The beam current, used in calculation of the irradiation dose, was measured using a Faraday cup. The beam current for all the irradiation experiments in this work was 30.8 pA. The experiments were conducted in a following sequence. First, the back gated graphene devices were irradiated with a certain dose of electrons. Second, the irradiated graphene devices were examined using micro-Raman spectroscopy to detect any changes with the Raman signatures of graphene. Third, the current-voltage (I-V) characteristics were measured to examine the changes of electrical properties. After I-V data were collected, the irradiation dose was increased and all steps repeated.

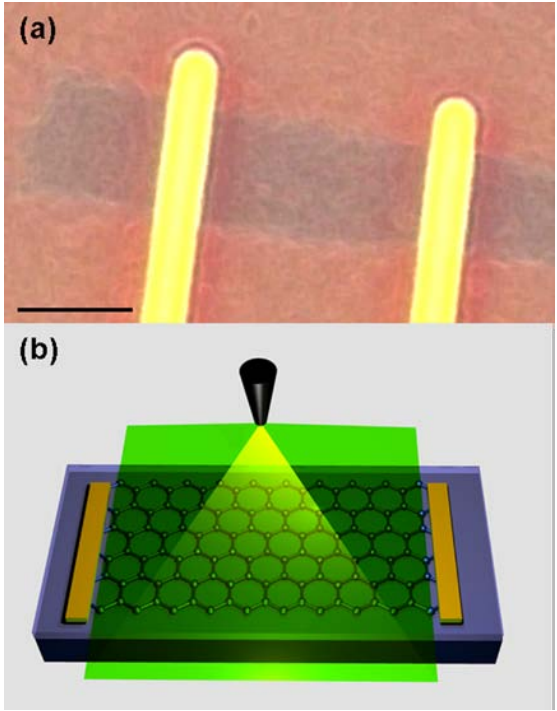

Fig. 1. (a) Optical image of a typical graphene device used in this work. The contrast is enhanced. The dark blue region is graphene. The metal electrodes are source and drain contacts, and heavily doped silicon wafer is used as a back gate. The scale bar is 2 μm. (b) Schematic of the irradiation by the electron beam. The green rectangular region is the irradiation area, which covers graphene between the source and drain while excludes two electrodes to avoid possible changes of the contact resistance due to irradiation.

The electron beam irradiation was performed inside the SEM vacuum chamber with a low pressure ($10^{-7}$ Torr) whereas the Raman spectroscopy and electrical measurements were carried out at ambient conditions. We used a Reinshaw InVia micro-Raman spectrometer system with the laser wavelength of 488 nm. The electrical measurements were performed with an Agilent 4142B instrument. Fig. 1 (a) shows an optical image of a typical SLG graphene device. In Fig. 1(b) illustrates the irradiation process showing the exposed and shielded regions of the device under test. The devices and irradiation process were intentionally designed in such a way that only graphene channel is exposed to the e-beam while the metal contacts are not irradiated. The latter allowed us to avoid any possible changes in metal contact resistance after the irradiation. We tested three SLG and three BLG devices.

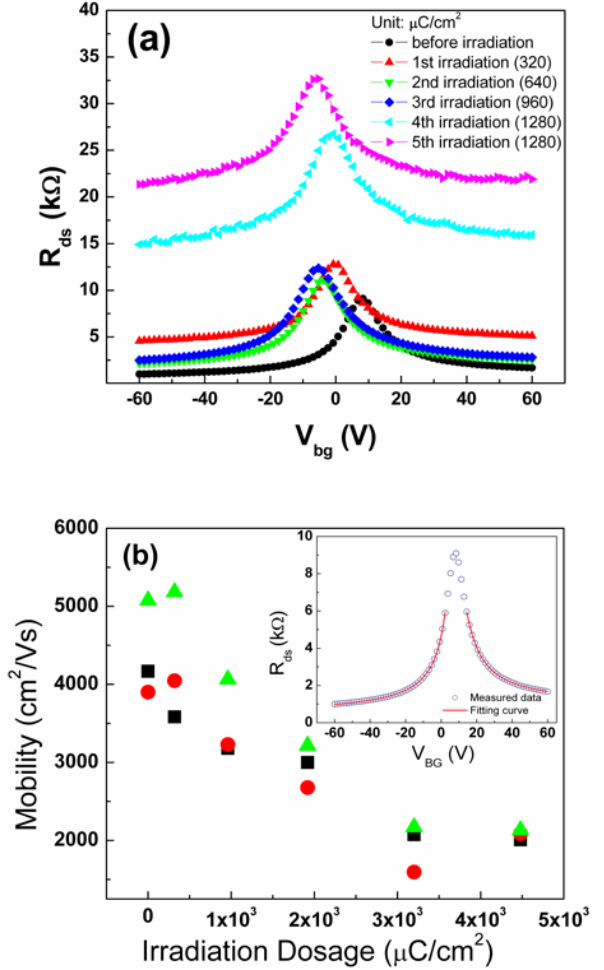

Fig 2. (a) Evolution of the transfer characteristics of SLG with increasing irradiation dose. The electrical resistance of SLG devices was measured after each irradiation step. The irradiation dose is indicated in the legend. (b) Charge carrier mobility as a function of the irradiation dose for three SLG devices, represents by red, green and black data points, respectively. Note a nearly linear decrease of the mobility with the irradiation dose. The inset shows the measured and fitted electrical resistance as a function of the back gate for one of the devices.

### III. RESULTS AND DISCUSSION

#### A. Single layer graphene devices

We started by measuring the electrical resistance between the source and drain as a function of the applied gate bias. Fig. 2 (a) shows the evolution of the electrical characteristics of SLG device after each irradiation step. The electron irradiation dose for each step is indicated in the figure's legend. As one can see, the ambipolar property of graphene is preserved after irradiation within the examined dosage range. The observed up shift of the curves indicates increasing resistivity of graphene over a wide range of carrier concentration. The increase is especially pronounced after the 4th step with a higher irradiation dose (1280 µC/cm²).

In order to analyze the results and rule out the role of the contact resistance we used the following equation to fit our resistance data [19, 20]

$$R_{ds} = R_{Cont} + \frac{L}{W}\left(\frac{1}{e\mu(\sqrt{n_0^2 + n_{BG}^2})}\right) \quad (1)$$

where $R_{Cont}$ is the contact resistance, µ is the mobility, $e$ is the elementary charge, $L$ and $W$ are the length and width of the channel, respectively. In Eq. (1) $n_0$ is the background charge concentration due to random electron – hole puddles [14] and $n_{BG}$ is the charge induced by gate bias calculate from the equation

$$n_{BG} = \frac{C_{BG}|V_{BG} - V_{BG,min}|}{e}, \quad (2)$$

where $C_{BG}$ is the gate capacitance per unit area taken to be 0.115 mF for 300 nm $SiO_2$ substrate.

The inset to Fig. 2 (b) shows the result of the fitting with Eqs. (1-2) of the data for SLG device before e-beam irradiation. Note that the fitting dose not cover the interval close to the charge neutrality point because this region is characterized by a large uncertainty in the data. The fitting was conducted separately for the negative and positive gate bias regions. For simplicity, we consider the fitting results from the p-type branch. The fitting gives the contact resistance of 446 Ω, the initial mobility µ=5075 cm²/Vs, and the charge impurity concentration of $2.13 \times 10^{11}$ cm⁻², which are very close to the typical values for clean graphene samples [21]. During the experiments the irradiated regions excluded the contacts. For this reason, the contact resistance should not change during the measurements and we can estimate the resistance of the irradiated graphene channels by subtracting the contact resistance from the total resistance. To fit our results for irradiated graphene devices we modified Eq. (1) by adding the term $R_{Irrd,}=(L/W)\rho_{Irrd}$, which is the resistance increment induced by e-beam irradiation. Fig. 2 (b) shows the evolution of the mobility due to e-beam irradiation for three SLG devices. We note that the mobility decreases almost linearly and drops by 50~60% over the examined irradiation dose.

We carefully examined the Raman spectrum of the graphene devices after each irradiation step. One can see from Fig. 3 (a) that the pristine graphene has typical signatures of SLG: symmetric and sharp 2D band (~2700 cm⁻¹) and large I(2D)/I(G) ratio. The absent or undetectably small D peak at 1350 cm⁻¹ indicates the defect-free high-quality graphene. The disorder D peak appears after the electron beam irradiation. Initially the intensity of the D grows with increasing dosage after each irradiation step. The trend reverses after the irradiation dose reaches a certain level. We used the intensity ratio I(D)/I(G) to characterize the relative strength of the D peak [13, 22]. The ratio I(D)/I(G) reveals a clear and reproducible non-monotonic dependence on the irradiation dose (see Fig. 3 (b)). This behavior was observed in all devices in our experiments. It is consistent with our earlier studies [13]. A similar trend was reported for graphite where the ratio I(D)/I(G) was also increasing with the irradiation dose. Such dependence was attributed to the crystal structure change from crystalline to nanocrystalline and then to amorphous form [22]. The bond breaking in such cases is likely chemically induced since the electron energy is not sufficient for the ballistic knock out of the carbon atoms [13]. Other factors contributing to the growth of the disorder D band can be contaminant molecules or water vapor, which dissolve under irradiation and may form bonds with the carbon atoms of the graphene lattice.

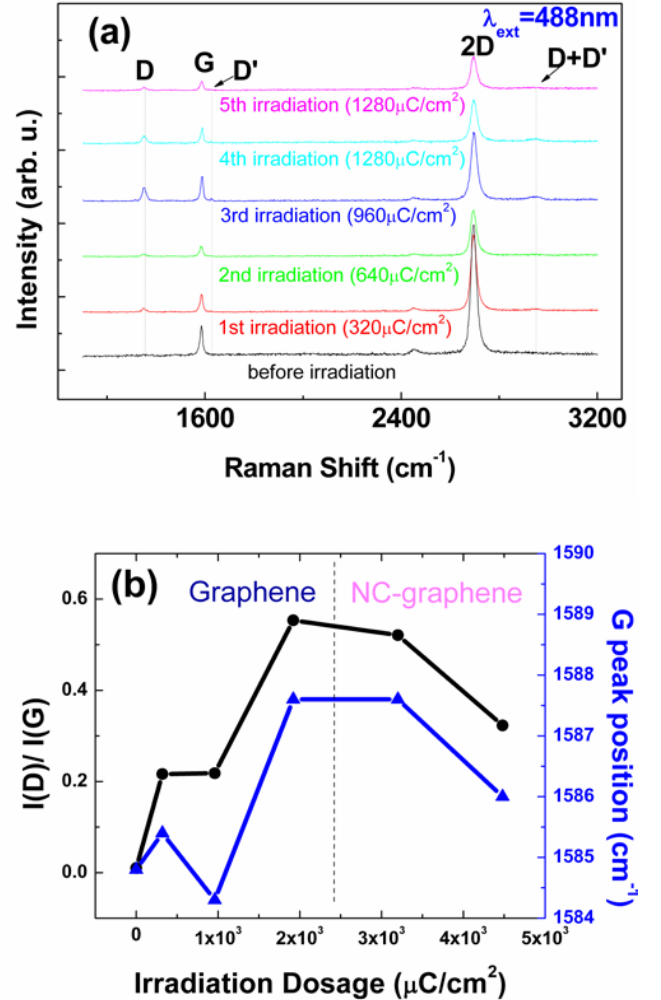

Fig. 3. (a) Evolution of Raman spectrum of SLG with increasing irradiation dose. The spectrum of pristine graphene before irradiation does not reveal the disorder band. A pronounced disorder D peak near ~1350 cm⁻¹ appears after irradiation. Another D' peak (~1620 cm⁻¹) and higher order harmonic D+D' (~2950 cm⁻¹) are also induced by irradiation. (b) The ratio I(D)/I(G) initially increases with the irradiation dose but starts to decrease after the 3rd irradiation step (black curve). The G peak position also reveals a non-monotonic dependence with the irradiation dose following a similar trend as the I(D)/I(G) ratio.

The change in the G peak position under the electron beam irradiation is shown in Fig. 3 (b). The G peak position shifts to higher wave numbers with increasing irradiation dose (with exception for the 2$^{nd}$ step). But after certain dose (step four) the peak position starts to move to the lower wave numbers. A similar trend was also observed in graphite [22]. It is reasonable to believe that e-beam irradiation leads to disorder in graphene's crystal lattice via formation of defects and sp$^3$ bonds.

In addition to D peak we also observed the appearance of other peaks in Raman spectrum of irradiated graphene. The peak at ~1620 cm$^{-1}$, referred to as D', was detected after the second step of irradiation. This peak was attributed to the intra-valley double-resonance process in the presence of defects [7]. The electron beam irradiation also results in the appearance of the D + D' peak around 2950 cm$^{-1}$. This peak, unlike the 2D and 2D' bands, is due to a combination of two phonons with different momentum and requires defects for its activation. A slight broadening of 2D band and decrease of the I(2D)/I(G) ratio were also observed. The decrease of the I(2D)/I(G) ratio was previously attributed to increasing concentration of charged defects or impurities [23]. Our electrical measurements are consistent with this interpretation indicating a growing density of the charged impurities with increasing irradiation dose (see inset to Fig. 4).

Fig. 4 shows evolution of the resistivity near the charge neutrality point with the irradiation dose. One can see a clear trend of increasing $\rho_{max}$ with the irradiation dose. Since the contacts were not irradiated during the experiment, the overall increase of device resistance is due to the increasing resistivity of the irradiated graphene. This can be understood by the induced defects that create an increasing number of scattering centers in the graphene lattice. Note that the $\rho_{max}$ increases by a factor of ~ 3 to 7 for SLG devices.

We also found that the irradiation induced changes in the properties of SLG are reversible to some degree. The IV characteristics can be at least partially recovered by annealing or storing the devices over a long period of time in a vacuum box. The annealing may help to repair the bonds and clean the surface from the organic residues while keeping devices in vacuum may lead to the loss of the irradiation induced charge. The latter suggests that the e-beam irradiation results in creation of the charged defects, which are more efficient in carrier scattering than neutral defects.

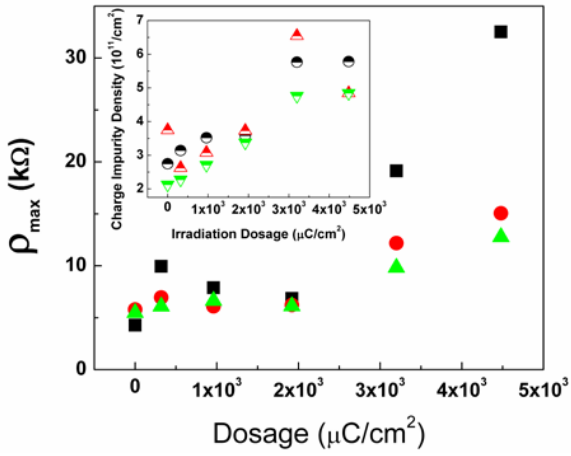

Fig. 4. Evolution of SLG resistivity with irradiation dose. The inset shows the effect of e-beam irradiation on the charge density for three SLG devices, represents by red, green and black data points, respectively.

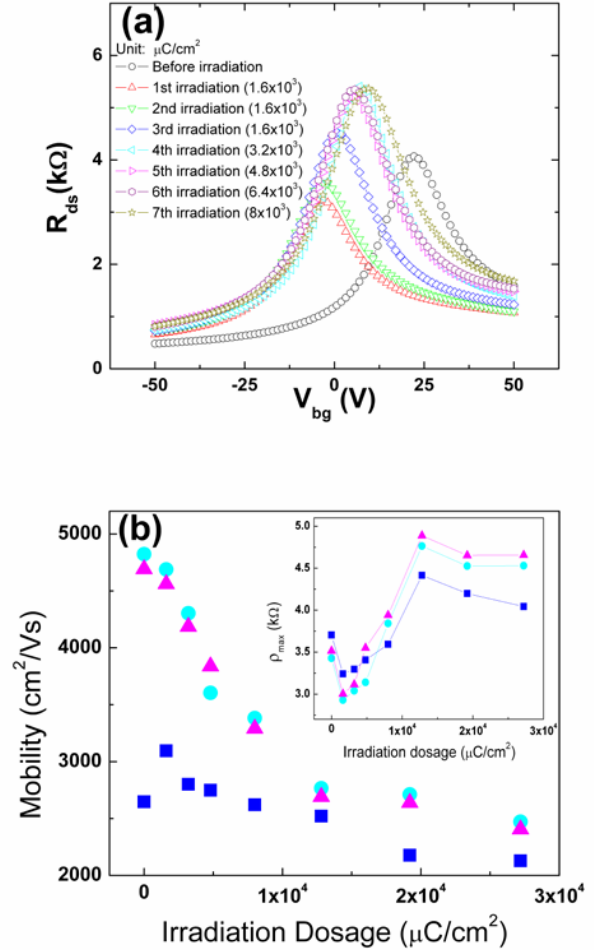

Fig. 5. (a) Evolution of the transfer characteristics of BLG with increasing irradiation dose. The irradiation dose after each step is indicated in the legend. (b) Carrier mobility of BLG devices as a function of the irradiation dose for three BLG devices, shown by pink, cyan and blue data points, respectively. Note that the for two devices with higher mobility the dependence has a turning point at the dose of about 12000 μC/cm$^2$ but for the device with lower mobility the decrease is approximately linear. The inset shows the electrical resistivity as a function of the irradiation dose.

### B. Bilayer graphene devices

In order to compare SLG with BLG under e-beam irradiation we conducted the same experiments with the back gated BLG devices. The only difference was a higher dose of irradiation for BLG than for SLG. The first step was 1600 μC/cm$^2$ compared to 320 μC/cm$^2$ in the first step for SLG. We expected that a larger dose would be required for BLG from the analogy with the multi-wall carbon nanotubes (CNTs), which were found to be less susceptible to e-beam irradiation than the single-wall CNTs [8]. We again used Raman spectroscopy to monitor the evolution of the material properties revealed by I-V measurements.

We observed substantially different irradiation induced effects in BLG as compared to SLG devices. Fig. 5 (a) shows evolution of the transfer characteristics for a typical BLG device with increasing irradiation dose. The total electron irradiation dose shown for BLG is 27200 $\mu$C/cm$^2$ while that for SLG is only 4480 $\mu$C/cm$^2$. In Fig. 5 (b) we present the effect of irradiation on the charge carrier drift mobility in BLG devices. One can see that the overall trend is similar to the SLG case but the mobility decrease rate is quite different. Our data indicate that the BLG is much less susceptible to e-beam irradiation than SLG. Indeed, if we look at the irradiation dose below 4480 $\mu$C/cm$^2$ we see that the mobility drop is smaller than 25% for BLG compared with ~50-60% drop for SLG. At the irradiation dose above 12000 $\mu$C/cm$^2$, the mobility decrease rate also reduces for the two high-mobility devices but for low-mobility devices the mobility decrease rate is roughly constant within the examined range. This is a similar behavior to the one reveled by SLG devices but requires much higher irradiation doses to be observed.

The resistivity $\rho_{max}$ increases by a factor of ~1.6 over the entire range for BLG devices as seen in the inset to Fig. 5 (b). Up to the dose of ~4480 $\mu$C/cm$^2$, $\rho_{max}$ of BLG changes only by ~14% compared to ~300-700% in the case of SLG. This difference is reflected by the I(D)/I(G) ratio in the Raman spectra for SLG and BLG.

The inset to Fig. 6 shows the Raman spectrum of a typical BLG device after several e-beam irradiation steps. Unlike in SLG the disorder induced Raman D peak in BGL does not reveal a pronounced growth with irradiation dose even over a much larger dose range. No detectable D' or D+D' peaks appear in the Raman spectrum of BLG. The absence of these peaks suggests e-beam irradiation over the examined dose range create limited amount of defects in BLG. Fig. 6 shows a comparison of the I(D)/I(G) ratio for two BLG with two SLG devices. The pristine BLG and SLG before irradiation have very small and comparable value of I(D)/I(G). The I(D)/I(G) ratio grows very fast in SLG devices with each irradiation step while it increases very slowly in BLG even over a wider irradiation dose range. This difference of I(D)/I(G) behavior in BLG and SLG is consistent with the different behavior of $\rho_{max}$ in BLG and SLG devices. Similar conclusions were made about the D peak induced by hydrogenation [7, 8]. The authors concluded that it is much harder to induce the disorder D peak in BLG than in SLG [7, 8]. A pronounced D peak in the Raman spectrum of BLG can be induced only using higher dose of e-beam irradiation [13, 14].

Our results suggest that BLG devices can perform better than SLG devices in applications which require radiation hardness. It has to be taken into account that irradiation may not only decrease the carrier mobility and electrical conductivity but also affect the excess noise level in such devices. The low level of 1/$f$ noise is essential for the proposed graphene applications in communication systems [24]. It was recently shown that graphene devices reveal a rather low level of 1/$f$ noise [25 – 27] but can degrade as a result of aging and environmental exposure [28]. The e-beam irradiation may lead to further increase in the noise level in graphene devices. For this reason, special protective cap layers may be required for communication and radiation-hard applications.

From the other side, the e-beam irradiation may lead to a new method of *defect engineering* of graphene physical properties. The controlled exposure of graphene layers to electron beams can be used to convert certain regions to the highly resistive or electrically insulating areas needed for fabrication of graphene circuits. Irradiation can also be used to reduce the intrinsically high thermal conductivity [4-6] to the very low values required for the proposed thermoelectric applications of graphene [29]. It is known from the theory of heat conduction in graphene that the lattice thermal conductivity can be strongly reduced by the defects and disorder [30-32]. The small-dose irradiation can become an effective tool for shifting the position of the minimum conduction point or inducing the carrier "transport gap."

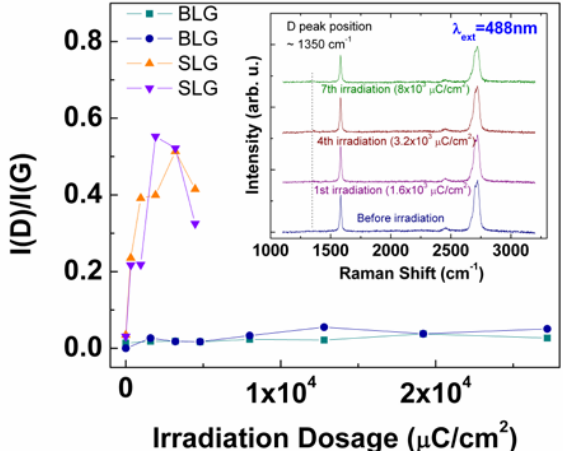

Fig. 6. Evolution of Raman spectrum of BLG with increasing irradiation dose. The examined BLG samples do not reveal either a prominent disorder D peak, or D'. The I(D)/I(G) intensity ratio is very small as compared with that in SLG. The data suggest that BLG graphene is much less susceptible to the electron beam irradiation than SLG.

## IV. CONCLUSIONS

We carried out detail investigation of the electrical and Raman spectroscopic characteristics of graphene and bilayer graphene under the electron beam irradiation. It was shown that the single layer graphene is much more susceptible to e-beam irradiation than bilayer graphene. The appearance of the disorder induced D peak in graphene Raman spectrum suggests that e-beam irradiation induce defects in graphene lattice. The mobility and electrical resistivity of graphene can be varied by the e-beam irradiation over a wide range of values. The obtained results may lead to a new method of defect engineering of graphene properties. The results also have important implications for fabrication of graphene nanodevices, which involve scanning electron microscopy and electron beam lithography.